# Designing high-transmission and wide angle all-dielectric flat metasurfaces at telecom wavelengths

**N. Vasilantonakis[*], J. Scheuer, and A. Boag**

*Department of Physical Electronics, Tel Aviv University, Tel Aviv 69978, Israel*

nvasilan@iesl.forth.gr

**Abstract:** Recent advances in holography, wireless sensing and light fidelity technologies have resulted in the need for antennas that can support highly efficient beam directivity for a broad angular range. Transmitarrays have been shown to be promising candidates for the determination of such structures. Here, we present a comprehensive methodology for the design of subwavelength all-dielectric flat metasurfaces with high transmission and deflection range. The metasurfaces are CMOS compatible and can be fabricated using conventional processing of silicon-on-insulator technology. The optimized structures exhibit up to 95% efficiency for a wide-angle deflection range from 20° to 60° in air, at operating wavelengths near 1550 nm. Furthermore, we propose the potential multi-wavelength and muti-mode excitation capabilities of the designed supercell. This is achieved by shifting the resonance of each post of the metasurface to the wavelength of interest opening up opportunities for wavelength multiplexing with a single structure. Our results facilitate the realization of broad angle flat deflectors and holographic devices but also indicate that today's bulky and expensive light fidelity optical routers could be substituted by ultra-small chips that are inexpensive to fabricate and can be commercially implemented.

## 1. Introduction

Over the last years there has been a remarkable research progress in the field of optical metamaterials which comprise subwavelength size features allowing the control of light-matter interaction. As a result, numerous functionalities have been proposed such as negative refractive index lenses [1], optical nonlinearity [2-4], cloaking [5-7], and optical magnetism [8, 9].

An emerging branch of metamaterials are the metasurfaces [10], which are planar arrangements of subwavelength size elements possibly embedded in layered media and can be seen as a two-dimensional (2D) case of metamaterials. The major advantages of metasurfaces as compared to conventional metamaterials, are the simpler and cheaper fabrication process and smaller losses. Their planar nature has found a plethora of applications in polarization conversion [11], beam deflection [12], 2D [13] and 3D [14] holography, and high numerical aperture metalenses [15].

The elements used in plasmonic metasurfaces can produce the required $2\pi$-phase manipulation using two different methods. The first method is via the excitation of two different resonances with a $\pi$-phase range for each, such as the V-shaped nano-antennas which can excite symmetric and asymmetric modes by tweaking the lengths and orientation of the nano-antennas [16]. The second method is through the spatial variation of geometric orientation of the plasmonic metasurface due to the Pancharatnam–Berry phenomenon [17]. Nevertheless, the intrinsic high loss of metals results in poor efficiencies for plasmonic metasurfaces specifically for transmitarrays [18].

Consequently, there has been significant research effort aimed at designing dielectric metasurfaces exhibiting negligible losses in the visible and infra-red ranges. When the refractive index of the dielectric building block is high, typically higher than 2, the nanoparticles may exhibit Mie-type resonances similar to plasmonic counterparts [19]. The building block is usually Silicon (Si) due its high refractive index and CMOS compatibility [20]. As an example, figures 1(a)-(d) show the cross-section of a unit cell with amorphous Si post with refractive index of about 3.5 [21] near telecom frequencies. Substrate is fused silica ($SiO_2$) with an index of 1.44 and the embedded medium is SU8 with an index of 1.58. The $SiO_2$ post exists for fabrication reasons. The thickness of $SiO_2$ and Si posts is 400 nm and 250 nm, respectively, while the unit cell is square with 770 nm length. The electric and magnetic field distributions depict an electric resonant behaviour (electric dipole) for a wavelength of 1500 nm, with the E-field accumulated at the centre of the Si post and a vortex-like H-field pattern (figures 1(a),(c)). Conversely, the magnetic resonance (magnetic dipole) has the magnetic field at the centre of the post and a vortex-like electric field around it (figures 1(b),(d)). When the two resonances overlap not only spatially but also in frequency the metasurface acts as Huygens' surface and near unity transmittance is feasible [22]. Figure 1(e) represents the transmittance dispersion at various radii revealing the overlapping of the two resonances close to 1500 nm, while figure 1(f) shows the corresponding $2\pi$-phase change. To observe this more clearly, figure 1(g) demonstrates the cross-section of transmittance and phase at 1500 nm. The reason of such high transmittance is the destructive interference of backscattered waves and constructive interference of transmitted ones when the electric and magnetic dipoles overlap, as can be seen from figure 1(h).

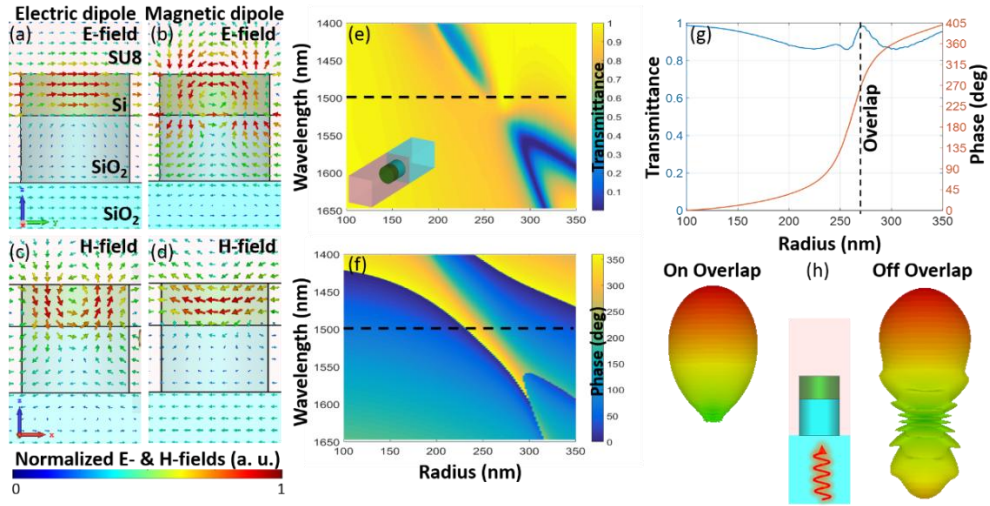

**Fig. 1**. (a,b) Electric and (c,d) magnetic field distributions of the electric and magnetic dipole resonances inside a high-index (Si) cylindrical post, with periodic boundaries considered, at 1500 nm. (e) Transmittance dispersion at various radii showing the near unity transmittance on the overlapping of the electric and magnetic dipoles near 1500

nm, exhibiting a 2π-phase change (f). (g) Cross-section of transmittance and phase at 1500 nm. (h) Far-field distribution of a single post on and off the overlapping of the electric and magnetic dipoles.

Since Huygens' metasurfaces allow high transmittance and arbitrary phase profiles, there have been a plethora of applications based on this concept such as highly efficient beam deflectors [13, 23] and holograms [13, 24]. Another very promising application of dielectric metasurfaces is their usage as a building block for flat and ultra-thin optical components for light fidelity (Li-Fi) systems [25]. However, as Li-Fi is an evolving field, the commercially available optical routers are bulky and expensive making them less attractive for deployment in offices and homes. A potential solution to this problem is the replacement of the current bulky and costly optical routers with ones based on chips comprising dielectric nano-antennas which are small, thin, cheap to fabricate, and have an area of only a few cm$^2$.

Recently, a new design approach for the realization of optimized all-dielectric metasurfaces was proposed [26]. In this work we extend this concept and present a comprehensive methodology for the design of subwavelength all-dielectric metasurfaces with high efficiency and controlled deflection angle at telecom wavelengths. More precisely, section 2 shows the general structure setup along with robustness analysis for the determination of the most crucial parameters that affect the overall response. Section 3 depicts the design and optimization process for the realization of single beam deflectors at various transmitted angles. Since more optimized parameters are considered here, efficiencies of up to 95% are achieved covering a broad far-field deflection angle range from 20° to 60° in air. These efficiencies are considerably higher than other similar works [13, 27, 28]. Furthermore in section 4, we demonstrate the potential multi-wavelength and multi-mode excitation capabilities of the designed metasurface by tuning the spectral position of the modes under consideration for a broad wavelength range around telecom frequencies. Finally, section 5 summarizes the main results.

## 2. General structure setup and robustness analysis

The general design of the structure is depicted in figure 2. A plane wave is excited in air and reaches the substrate, SiO$_2$, at normal incidence. The thickness of substrate, $d_{sub}$, is around 2 μm and thus considered infinite in calculations. The dielectric metasurface consists of a number of cylindrical posts, $N$, that were varied from 2 to 6; higher number of posts give lower deflection angles that were out of the scope of the present study. The length of supercell in x- and y-directions is $L_x$ and $L_y$, respectively. The radius of each post is denoted $R_1,…,R_6$ with centre coordinates $(X_1, Y_1),…,(X_6, Y_6)$. The origin of coordinate system is set to be in the middle of the supercell, which is at $L_x/2$ and $L_y/2$. The embedding medium, SU8, has a thickness of $d_{SU8} = d_{SiO2} + d_{Si} + d_{over}$, where $d_{SiO2}$, $d_{Si}$, and $d_{over}$ are the thicknesses of the SiO$_2$ posts, Si posts, and the overlayer, respectively. Finally, an anti-reflection coating with refractive index of 1.28 and thickness $d_{AR}$, is added between SU8 and air as a matching layer to minimize reflection.

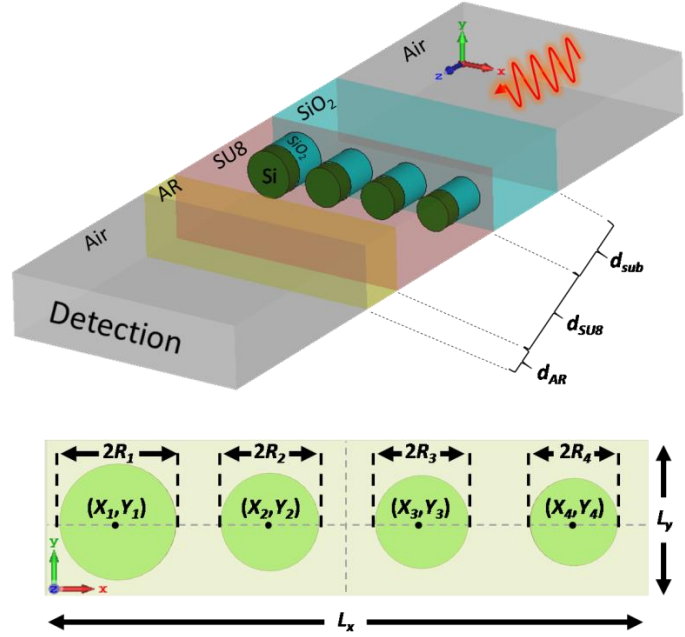

**Fig. 2**. Supercell of the simulated structure showing the parameters that were optimized. The bottom image depicts the top view of the supercell. Note that in simulations the number of posts inside the supercell varied from 2 to 6.

Based on the Generalized Snell's Law (GSL) [29], it is possible to achieve non-zero deflection angle, even if the incidence angle is zero, by adding a subwavelength phase change on the interface. This can be achieved via the manipulation of the Si posts on the metasurface. As a result, the main objectives are to control the deflection angle and, simultaneously, maximize the efficiency of the transmitted Floquet mode(s) at a fixed operating wavelength. To achieve this, structure optimization is necessary. At first, the Floquet modes of a given periodic structure were computed via the Rigorous Coupled Wave Analysis (RCWA) method [30, 31]. Then optimization was performed through Genetic Algorithm (GA). Lastly, the optimized structure was simulated and compared with CST commercial software to ensure validity.

At first, robustness analysis is performed in order to inspect the impact in efficiency for the case where the fabricated structures deviate from the optimized geometrical parameters. Figures 3(a),(b) demonstrate the stability of the optimized geometry for 2 posts, while figures 3(c),(d) - for 3 posts, at a fixed wavelength of 1550 nm. From figure 3(a), we observe that the radii ($R_1$, $R_2$) and thickness ($d_{Si}$) variations of the Si posts have the most dramatic effect on the efficiency causing more than 50% decrease for a ±10 nm change from the optimized parameters. Similarly, figure 3(c) shows that radii ($R_1$, $R_2$, $R_3$) and thickness ($d_{Si}$) of Si posts have the greatest impact on the efficiency that reaches up to 40% decrease from the optimized scenario. On the contrary, the thicknesses of $SiO_2$ posts ($d_{SiO2}$), overlayer ($d_{over}$) and anti-reflection coating ($d_{AR}$) have negligible impact with less than 1% efficiency decrease. Figures 3(b),(d) demonstrate that the size of the supercell ($L_x$, $L_y$) and x-coordinates of the posts ($X_1$, $X_2$, $X_3$) are also the dominating factors producing up to 12% decrease in efficiency at ±10 nm shift from reference parameters, while y-coordinates have a vanishing small impact (<< 1%). We thus conclude that the radii and thickness of Si posts are the most critical parameters that need to be carefully controlled in fabrication, for instance, through the silicon-on-insulator (SOI) method. The reason for this dependence is the high refractive index of Si which leads to high field localization inside the Si posts mostly affecting the transmission efficiency. The robustness analysis was repeated for higher numbers of posts confirming the strong dependence of the overall response on the Si post dimensions (not shown).

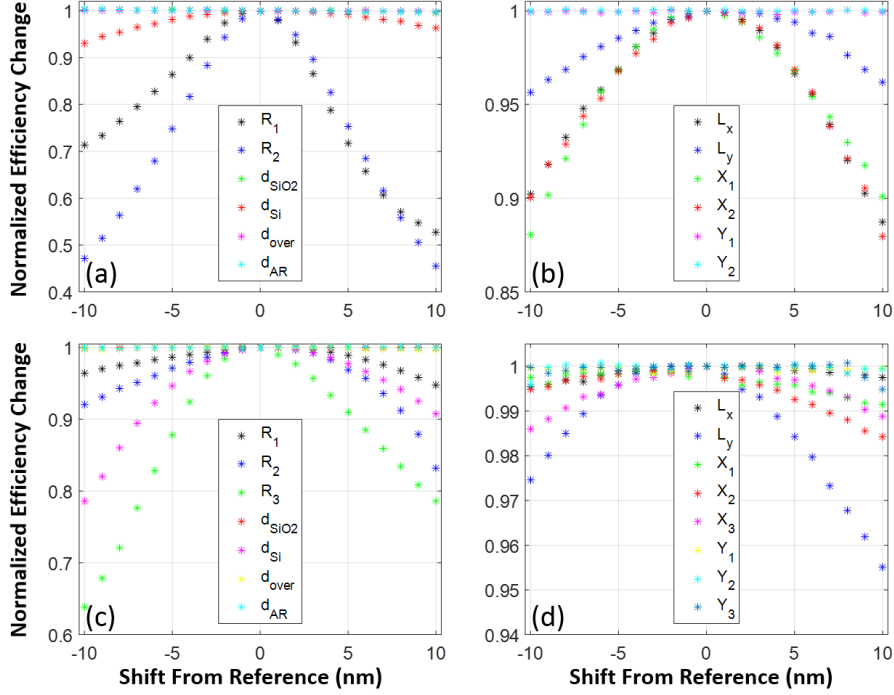

**Fig. 3**. Robustness analysis of the optimized parameters at a fixed wavelength of 1550 nm, for 2 (a,b) and 3 (c,d) posts in the supercell. In both cases robustness is mainly affected by the thickness and radii of Si posts.

## 3. Design and optimization of single beam deflectors

This section demonstrates the main results for the design of single beam deflectors at wavelengths close to 1550 nm. Table 1 has the cumulative values of the optimized parameters for all the examined supercells from $N = 2$ to $N = 6$. As mentioned in the previous section, the $y$-coordinates produce negligible change in efficiency and are excluded from optimization. Figures 4(a) and 4(b) compare the transmittance of the +1 Floquet mode between optimized and non-optimized geometries, respectively. It is clear that for all the optimized geometries, $N = 2$ to 6, the resonance falls close to the operating wavelength maximizing the transmittance (figure 4(a)), while for the non-optimized case the resonance is spectrally far giving poor performance. For almost all cases the optimized transmittance is higher than 80% and reaches up to 95% for $N = 4$. The bandwidth, defined as the spectral range which the efficiency is more than 90% of the highest value, is 12 nm, 25 nm, 4 nm, 20 nm and 18 nm as $N$ varies from 2 to 6 respectively. Figures 4(c) and 4(d) depict the electric field distributions of the cross-section of supercells with $N = 2$ and $N = 6$, respectively. These 2 extreme cases show how the deflection angle varies from high to low values. To observe this more clearly, figure 4(e) demonstrates the far-field optical response for all the aforementioned geometries. For low number of posts ($N = 2$) the deflection angle is 60° and as we gradually increase $N$ the angle decreases to 20°. It should be noted that the angle variation is not linear versus $N$. As figure 4(e) shows, there is a 23° gap between $N = 2$ and $N = 3$. This can be compensated by changing the lattice of the supercell; for example switching from square (figure 2) to hexagonal lattice (not shown). For the case of a single beam deflector that deflects light to negative angles, there are two possible ways: A) to design supercells with optimized transmittance of the -1 Floquet mode or B) to rotate the plane of metasurface by 180° since the angle of incidence is normal to the interface. Lastly, figure 4(f) summarizes and compares the optimized efficiencies for various deflection angles, computed using CST and RCWA codes. Both methods give similar results and efficiency ranges from 72% ($N = 2$) up to 95% ($N = 4$).

Table 1. Optimized parameters of the single beam deflector.

| $N$ | $d_{SiO2}$ (nm) | $d_{Si}$ (nm) | $d_{over}$ (nm) | $d_{AR}$ (nm) | $L_x$ (nm) | $L_y$ (nm) |
|---|---|---|---|---|---|---|
| 2 | 460 | 360 | 925 | 453 | 1782 | 841 |
| 3 | 760 | 293 | 1210 | 158 | 2558 | 758 |
| 4 | 888 | 303 | 1192 | 219 | 3096 | 714 |
| 5 | 713 | 310 | 1238 | 27 | 3983 | 705 |
| 6 | 245 | 358 | 1047 | 321 | 4564 | 648 |
| | $R_1$ (nm) | $R_2$ (nm) | $R_3$ (nm) | $R_4$ (nm) | $R_5$ (nm) | $R_6$ (nm) |
| 2 | 156 | 187 | - | - | - | - |
| 3 | 167 | 235 | 258 | - | - | - |
| 4 | 161 | 206 | 232 | 301 | - | - |
| 5 | 196 | 216 | 244 | 261 | 276 | - |
| 6 | 107 | 184 | 224 | 229 | 243 | 272 |
| | $X_1$ (nm) | $X_2$ (nm) | $X_3$ (nm) | $X_4$ (nm) | $X_5$ (nm) | $X_6$ (nm) |
| 2 | -327 | 318 | - | - | - | - |
| 3 | -672 | -31 | 732 | - | - | - |
| 4 | -993 | -430 | 476 | 1124 | - | - |
| 5 | -1294 | -796 | -22 | 721 | 1418 | - |
| 6 | -1814 | -1161 | -502 | 311 | 1107 | 1826 |

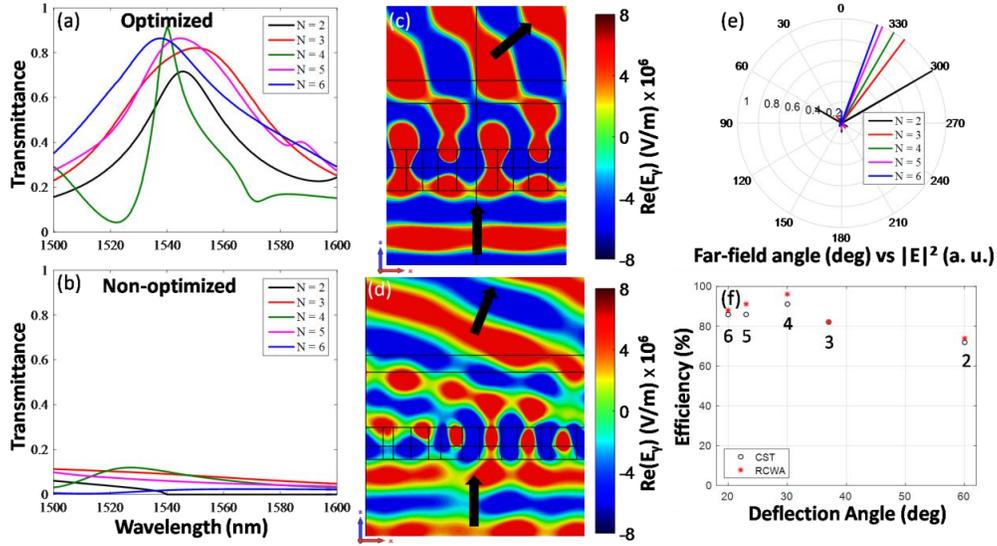

**Fig. 4**. Single beam deflector. (a) Optimized and (b) non-optimized transmittance for various numbers of posts (*N*) for wavelengths close to 1550 nm. (c) Electric field distribution for *N* = 2 and (d) *N* = 6 showing how the deflection angle greatly changes between the two cases. Arrows represent the direction of the incoming and deflected beam. (e) Far field diagram for different numbers of posts demonstrating the change of deflection angle from 20° (*N* = 6) to 60° (*N* = 2). (f) Optimized efficiency for various deflection angles computed by RCWA and CST. The number below the data points is the number of posts.

## 4. Design and optimization of multi-wavelength and multi-mode deflectors

As a next step, we investigate the multi-wavelength and multi-mode capabilities of the dielectric metasurfaces near telecom frequencies. Such component could prove useful for devices operating at multiple wavelengths or for wavelength division and spatial multiplexing, deflecting different data streams (at dissimilar wavelengths) to different locations. The simplest scenario is to make a supercell that consists of two posts and optimizing each post to have its resonance at a given frequency. Figure 5 presents the optimized transmittance diagrams for the +1 and -1 Floquet modes. The optimized parameters are $d_{SiO2}$ = 593 nm, $d_{Si}$ = 254 nm, $d_{over}$ = 631 nm, $d_{AR}$ = 301 nm, $L_x$ = 1758 nm, $L_y$ = 836 nm, $X_1$ = -287 nm, $X_2$ = 345 nm, $R_1$ = 160 nm, while $R_2$ ranged from 295 nm to 310 nm. The supercell is designed to have the +1(-1) Floquet modes at slightly longer(shorter) wavelengths than 1550 nm. The +1 mode exhibits an efficiency of 50% that remains approximately unchanged as it shifts from 1550 to 1600 nm having a bandwidth of approximately 27 nm (figure 5(a)). Similarly, the -1 mode has an efficiency of around 30% for a wavelength range spanning from 1500 nm to 1550 nm with a bandwidth close to 6 nm (figure 5(b)). The peaks of figure 5(a) and the large bandwidth resonances of figure 5(b) located at 1550-1600 nm are mostly from the contribution of the post with the greater radius ($R_2$), while the small bandwidth resonances of figure 5(b) are from the contribution of the post with the smaller radius ($R_1$). It should be emphasized that the efficiencies in both Floquet modes are significantly smaller compared to the efficiencies of figure 4(a). The reason is the extra optimization conditions that need to be satisfied simultaneously for the multi-wavelength/mode excitation to be feasible.

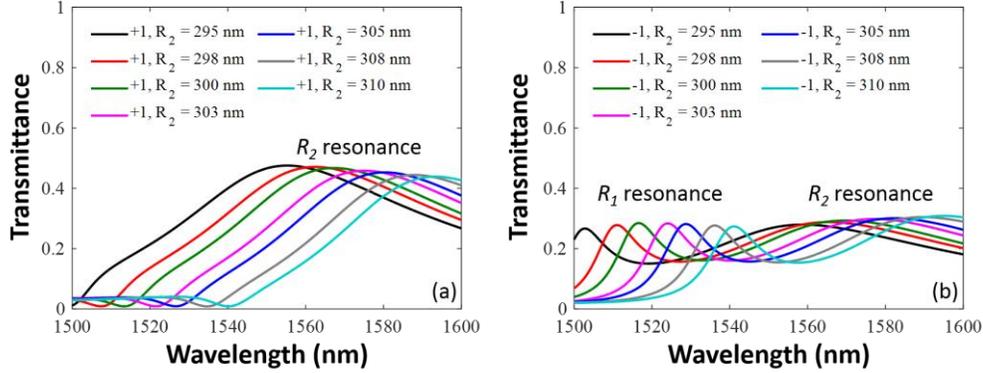

**Fig. 5**. Multi-wavelength and multi-mode deflector. Optimized transmittance diagrams for the +1 (a) and -1 (b) Floquet modes for various $R_2$. The graphs show how the efficiency can be tuned for different wavelengths retaining its amplitude at approximately 50% for the +1 and 30% for the -1 Floquet modes.

## 5. Conclusions

In summary, we have presented the design and optimization procedure of flat and subwavelength dielectric metasurfaces for highly efficient and wide angle beam deflectors. The metasurfaces exhibited efficiencies higher than 80% at an operating wavelength close to 1550 nm and for a broad far-field angular deflection ranging from 20° to 60° in air. The maximum efficiency achieved was 95% at 30° in air, which is significantly higher than other design methods. The robustness analysis showed that efficiency is more sensitive to size variations of the high-index (Si) medium due to the high field localization inside the Si-posts. For instance, a 10 nm change of Si radius from the optimized case, can have a 50% decrease in efficiency contrary to other parameters such as the thickness of the other layers, the length of supercell or the coordinates of the posts' centre, where the respective efficiency change is no more than 10%. Furthermore, we demonstrated how such structures can support multi-wavelength and multi-mode excitation with a controllable spectral position opening up opportunities for wavelength multiplexing near telecom frequencies. In such a case however, the efficiencies were more modest since there are additional optimization conditions that need to be satisfied with a single structure. This work paves a way for the realization of ultra-small CMOS compatible chips that are inexpensive to fabricate and can be commercially implemented for Li-Fi optical routers or for optical components in holography and photonic integrated circuits.

## Acknowledgements

The authors acknowledge the financial support of the XIN center of Tel Aviv and Tsinghua Universities.